\magnification=\magstep1
\hsize=16.6truecm
\vsize=22.2truecm
\baselineskip=5.truemm
\parskip=2.truemm plus 1.0truemm minus 0.5truemm

\input epsf

\font \eightrm=cmr8
\font \sasbf=cmssbx10                     
\font \bigssbf=cmssbx10 scaled \magstep1  
\font \hugssbf=cmssbx10 scaled \magstep2  

\def\subtitle#1{{\bigskip\centerline{\bigssbf #1}\smallskip}}

\def\wII{$w_{\rm{II}}$}
\def\xcut{x_{\rm{cut}}}
\def\Vst{V_{\rm{Step}}}
\def\Nst{N_{\rm{St}}}

\centerline{\hugssbf About the Significance of Quasinormal Modes}
\smallskip
\centerline{\hugssbf of Black Holes}

\medskip
\centerline{Hans-Peter Nollert}

\smallskip
\centerline{\it Theoretische Astrophysik, Computational Physics,
Universit\"at T\"ubingen}
\centerline{\it 72076 T\"ubingen, Germany}

\bigskip
\centerline{\sasbf Abstract}
\medskip
\centerline{\vbox{\hsize=0.8\hsize\eightrm
Quasinormal modes have played a prominent role in the discussion of
perturbations of black holes, and the question arises whether they are
as significant as normal modes are for self adjoint systems, such as
harmonic oscillators.  They can be significant in two ways: Individual
modes may dominate the time evolution of some perturbation, and a
whole set of them could be used to completely describe this time
evolution. It is known that quasinormal modes of black holes have
the first property, but not the second. It has recently been suggested
that a discontinuity in the underlying system would make the
corresponding set of quasinormal modes complete. We therefore turn the
Regge-Wheeler potential, which describes perturbations of
Schwarzschild black holes, into a series of step potentials, hoping to
obtain a set of quasinormal modes which
shows both of the above properties. This hope proves to be futile,
though: The resulting set of modes appears to be complete, 
but it does not contain any individual mode any more which is directly
obvious in the 
time evolution of initial data. Even worse: The quasinormal frequencies
obtained in this way seem to be extremely sensitive to very small
changes in the underlying potential. 
The question arises whether - and how - it is possible to make any
definite statements about the 
significance of quasinormal modes of black holes at all, and 
whether it could be possible to obtain a set of quasinormal modes
with the desired properties in another way.
}}

\medskip

\medskip\noindent
Paper: gr-qc/9602032

\medskip
\subtitle{I. Introduction}
\def\Litw{1} \def\Litwii{2}
Quasinormal modes are single frequency modes 
dominating the time evolution of perturbations of systems which are 
subject to damping, either by internal dissipation or by radiating away
energy. Due to the damping, the frequency of a quasinormal mode must
be complex, its imaginary part being inversely proportional to
the typical damping time.
Examples are oscillations of stars damped by internal
friction. In general relativity, damping occurs even without friction,
since energy may be radiated away towards infinity by gravitational
waves. This leads to a new class of quasinormal modes of neutron stars
which are not present in the Newtonian case~[\Litw,\Litwii]. 
\def\Litqnmoc{3} \def\Litcompoc{4}
Recently it has also been suggested~[\Litqnmoc,\Litcompoc] to describe the
behavior of leaky optical cavities using a quasinormal mode formalism, rather
than studying the surrounding infinite system that the cavity is
embedded in.

\def\Litqnmbh{5} \def\Litnumqnm{6}
Even linearized perturbations of black holes exhibit quasinormal
modes~[\Litqnmbh], despite the absence of any oscillating fluid.
The characteristic modes of black holes have first
been found by numerical calculations~[\Litnumqnm], leading to  attempts to
develop a formalism to describe characteristic oscillations without
having to refer to a specific initial perturbation, based only on the
properties of the underlying system.

\def\LitRWeq{7}
The study of linearized perturbations of Schwarzschild black holes
leads to the Schr\"o\-din\-ger-like equation (e.g.~[\LitRWeq]) 

\def\EqSchr{1}
$$ \psi''(x) + \biggl(\omega^2 - V(x)\biggr) \psi(x) = 0  , \eqno{(\EqSchr)}
$$
where $\psi$ represents the wave function of the perturbation and 
$V(x)$ is the so-called Regge-Wheeler potential
\def\EqRW{2}
$$ V(x) = \left(1 - {1\over r}\right) 
                 \left({l(l+1)\over r^2} - {3\over r^3}\right)  .
\eqno{(\EqRW)} $$
$l$ originates from the expansion of the perturbation in terms
of spherical harmonics, 
and $r$ is considered a function of $x$ such that 
$x = r + \ln(r-1)$. All units are chosen such that $G = 1$ and $c =
1$, the radial coordinate is scaled such that the horizon is at $r =
1$, i.e. $2M = 1$, where $M$ is the mass of the black hole. 

Quasinormal modes are formally defined by those solutions of 
Eq.~(\EqSchr) which satisfy purely outgoing boundary conditions, i.e. 

\def\Eqbound{3}
$$ \psi(x) \buildrel x\rightarrow -\infty \over \longrightarrow 
              e^{+i\omega x} {\rm \qquad and \qquad }
\psi(x) \buildrel x\rightarrow +\infty \over \longrightarrow 
              e^{-i\omega x}      .                        \eqno{(\Eqbound)}
$$
We have chosen the time dependence to be $e^{i\omega t}$, quasinormal
frequencies of stable systems must therefore have a positive imaginary
part. 

\def\LitInfMany{8}
In the case of a Schwarzschild black hole, it is known [\LitInfMany] that there is
an infinite number of discrete frequencies $\omega$ which allow 
solutions of Eqs. (\EqSchr) -- (\Eqbound). However, 
the significance of these frequencies and their associated
modes is not immediately clear. In the case of normal modes, the
underlying mathematical problem is self adjoint, and the normal modes
form a complete set of solutions. Therefore, the behavior of the 
system can be analyzed completely in terms of its normal modes. 
There is no such theorem for quasinormal modes, since Eqs. (\EqSchr)
-- (\Eqbound) do not define a self adjoint problem.
The completeness of a set of normal modes also makes it possible to
define unambiguously what we mean by the excitation strength of a given
normal mode as a result of some initial data. Again, this
definition does not simply carry over to quasinormal modes, not even
to the fundamental ones which are obvious in the time evolved data.

\def\LitTails{9}
\def\LitHighMod{\LitTails}
Nevertheless, quasinormal modes are believed to play a significant
role for the study of perturbations of black holes. 
This view is supported by
the fact that the fundamental quasinormal frequency (the
one with the smallest imaginary part) is generally found to dominate the time 
evolution of some initial perturbation. Even a few of the more strongly damped
modes, which are not immediately obvious in the time evolved data, can be
extracted with suitable techniques [\LitHighMod].

However, the fundamental
quasinormal mode can only be an approximation of the actual time
evolved perturbation. 
An interesting question is whether the set of all quasinormal modes of
a black hole can be used, in analogy to a set of normal modes of a
self adjoint system, to
completely describe the behavior of a perturbation of the black hole, i.e.
the time evolution of some initial data for the perturbation equation.
Quasinormal modes cannot be complete in the usual sense, meaning that
at any given time, the solution can be represented, over all space, as
a sum of quasinormal modes.
Due to the boundary conditions (\Eqbound), solutions
corresponding to damped oscillations have to grow exponentially as $x
\rightarrow \pm \infty$. 

\def\LitTorsFib{10}
However, it may be possible to represent the time dependence of a
solution at a given point in space completely as a sum over
quasinormal modes. A model systems with this property has been
studied by Price and Husain~[\LitTorsFib].
In general, however, quasinormal modes do not form such a complete system.
This is related to Eq. (\EqSchr), together
with the boundary conditions (\Eqbound), not defining a selfadjoint problem.
Therefore, the question of the mathematical as
well as physical meaning of quasinormal modes and of their usefulness
in practical computations has to be studied for every 
system in particular. 

\def\Litcomppot{11}
Ching et al.~[\Litcomppot] have argued that a system which is
described by a Schr\"odinger-like equation such as Eq.~(\EqSchr) will
indeed have a complete set of quasinormal modes if the potential
$V(x)$ has a discontinuity either in the potential itself or in any of
its derivatives.

On the other hand, it is known that the quasinormal modes of a
Schwarzschild black hole (or of a neutron star) cannot be complete.
One reason is that the late time behavior of perturbations of a black
hole is first
dominated by damped oscillations ("quasinormal ringing"), but at very
late times it exhibits a power-law tail~[\LitTails]. Such a  tail cannot be
obtained by a sum of quasinormal modes. In addition, the set of quasinormal
frequencies of black holes does not contain frequencies with large
real parts, i.e. oscillations with very short periods. Therefore, any
initial data corresponding to such oscillations cannot be adequately
represented by quasinormal modes.

The argument of Ching et al.~[\Litcomppot] therefore implies that 
if the (smooth)
Regge-Wheeler potential is changed slightly such that a
discontinuity, however small, is introduced in the potential or any of its
derivatives, the set of quasinormal modes of the black hole 
will change from incomplete  to complete. On the one hand, this would
be very convenient, since it would allow the study of arbitrary
perturbations in terms of this expanded set of quasinormal modes, just
as the time dependent behavior of any self adjoint system can be
studied in terms of its normal modes. 
On the other hand, one would not expect
the physical response of the black hole, i.e. the time evolution of
some perturbation, to be significantly affected by a small change in
the potential. We are therefore facing a possible contradiction: The
actual time evolution of some initial data will probably  not be
affected significantly by this small change in the potential, while the
quasinormal mode spectrum, which is generally 
believed to represent crucial aspects of this time evolution, should
change drastically. 

In an attempt to shed some light on this apparent contradiction, we will
replace the Regge-Wheeler potential in Eq. (\EqSchr) by a step
potential. This is a  potential which is piecewise constant,
constructed in such a way that it approximates the original, continuous
Regge-Wheeler potential. 
Of course, we expect any given step potential to have a quasinormal
modes spectrum which is somewhat different from that of the smooth
Regge-Wheeler potential. Therefore, we construct a whole series of
such step potentials and study the behavior of
their quasinormal frequencies as the number of steps is increased,
i.e. as the step potentials are allowed to approximate the continuous
Regge-Wheeler potential better and better.

\vbox{

\subtitle{II. Procedure}
In the following, we will always use the Regge-Wheeler potential
(\EqRW) with $l = 2$. 
There are, of course, many different ways to construct step potentials
which eventually approximate a given smooth potential. We will present
the results for three such possibilities: 
}

{\parindent=2em

\item{(i)} The difference in potential between steps is (roughly)
  constant. In order to achieve this, the length of the steps is
  variable and depends essentially on the derivative of the smooth
  potential.
  
\item{(ii)} The length of the steps is constant, while the difference
  in height between the steps depends on the smooth potential.
  However, the length of the steps may be chosen differently to the left and
  to the right of the maximum of the potential if the potential is not
  symmetric around its maximum.

\item {(iii)} As in (i), but an exponential damping is applied to the
  smooth potential before the steps are constructed. 
\par}

\topinsert
\epsfxsize=\hsize \epsfbox{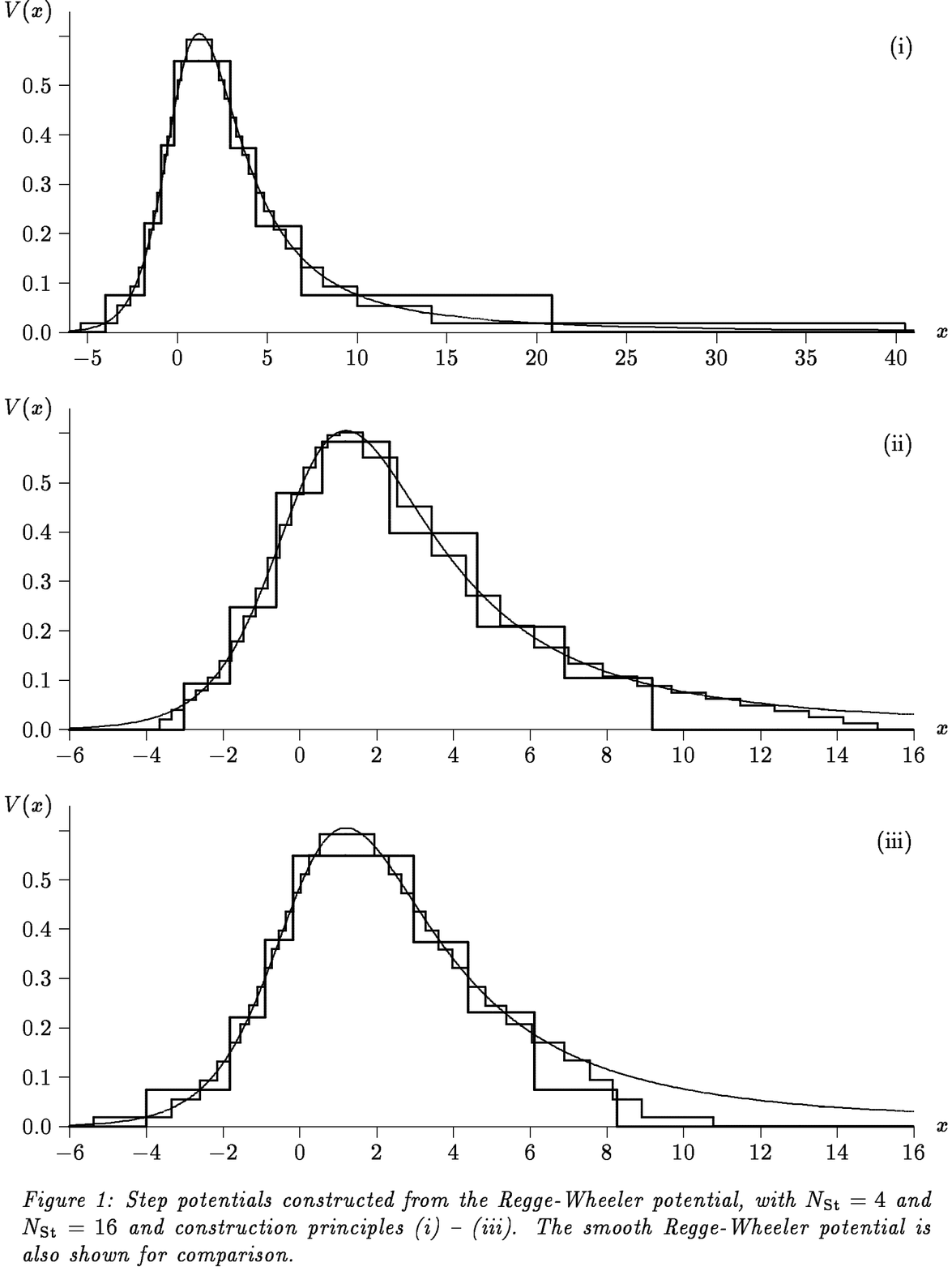}
\endinsert

\noindent
In all cases, the value of the step potential for a particular step is
generally determined by 

\def\Eqstep{4}
$$  \Vst(x_{n-1}\le x < x_{n}) = 
                 {1 \over x_{n} - x_{n-1}} \int_{x_{n-1}}^{x_n} V(x) dx
    \quad\hbox{for}\quad n = -\Nst+1 \dots \Nst,
\eqno{(\Eqstep)} $$
where $\Nst$ is the number of steps to either the left or the right of
the maximum of the potential.
For method (i), the position $x_{\Nst}$ of the last jump on the right 
is determined by

\def\Eqlast{5}
$$  \Vst(x_{\Nst-1}\le x < x_{\Nst}) = {V_{\rm max} \over 2\Nst} =
       {1 \over x_{\Nst} - x_{\Nst-1}} \int_{x_{\Nst-1}}^{\infty} V(x) dx
,  \eqno{(\Eqlast)}
$$ 
with a corresponding condition for the last step on the left. This
ensures that

\def\Eqint{6}
$$  \int_{-\infty}^{+\infty} \Vst(x) dx =
    \int_{-\infty}^{+\infty} V(x) dx         .
\eqno{(\Eqint)} $$
This condition is not satisfied by methods (ii) and (iii).

Figure 1 (i) -- (iii) shows the resulting step potentials for different
numbers of steps. Method (i) obviously leads to the last step
on the right of the potential's maximum becoming extremely long.
 This is due to the slow decay
of the Regge-Wheeler potential for $x \rightarrow \infty$: $V(x) \sim
1/x^2$ as $x \rightarrow \infty$, while $V(x) \sim \exp(x)$ as 
$x \rightarrow -\infty$, as can be verified from Eq.~(\EqRW). The last
step being so much longer compared to all other steps seems rather
artificial. In fact, with increasing $\Nst$, the last steps grows longer,
while the steps near the maximum of the potential become shorter, as
one would indeed expect for a refinement of the step potential.

Therefore, in method (iii) we have imposed an exponential damping on 
the potential before constructing the steps: 

\def\Eqexpcut{7}
$$ \tilde V(x) = \cases{{V(x) \phantom{/ \cosh^2({1 \over 2}(x-\xcut))}
                                                   \qquad x <= x_{cut}} \cr
                        {V(x) / \cosh^2({1 \over 2}(x-\xcut)) 
                                                   \qquad x > \xcut} \cr}
\eqno{(\Eqexpcut)} $$

Note that this leads to an exponential decrease with the same power as
for $x \rightarrow -\infty$. Of course, $\xcut$ itself has to grow
with increasing $\Nst$ if the step potential is supposed to approximate
the smooth potential arbitrarily well for $\Nst \rightarrow \infty$. We
use $\xcut := x_{\rm max} + 4 (\Nst/4)^{1/4}$, where
the maximum of the Regge-Wheeler potential is at $x_{\rm max} \simeq 1.1947$. 

A related problem occurs in method (ii): Here, the last step would
become much higher than the differences in height of the preceding
steps, again due to the slow decrease of the Regge-Wheeler potential
for $x \rightarrow \infty$. In order to avoid this, several steps
before the last are reduced in height such that the step height 
approaches 0 linearly. Again, this artificial decrease sets in at larger
values of $x$ if the number of steps increases.

The determination of the quasinormal modes of these step potentials is
quite simple: On the $n$th step, the solution is given by 
$\psi(x_{n-1}\le x < x_{n}) = A_n e^{ik_nx} + B_n e^{-ik_nx}$, 
with $k_n = \sqrt{\omega^2 - \Vst(x)}$, and
$A_n$, $B_n$ determined according to the boundary conditions. For a
given (complex) frequency $\omega$, we start
with $\psi_{+\infty} = e^{-i\omega x}$ to the right of the potential 
and, moving to the
left,  determine $A_n$ and $B_n$ at every step by the standard
procedure of matching $\psi(x)$ and $\psi'(x)$ at every step. If
$\psi(x) = \psi_{-\infty}(x) = e^{+i\omega x}$ after
reaching the left side of the support of the potential, i.e. if
$B_{-\Nst} = 0$, the frequency $\omega$ is a quasinormal frequency. 

\topinsert
\epsfxsize=\hsize \epsfbox{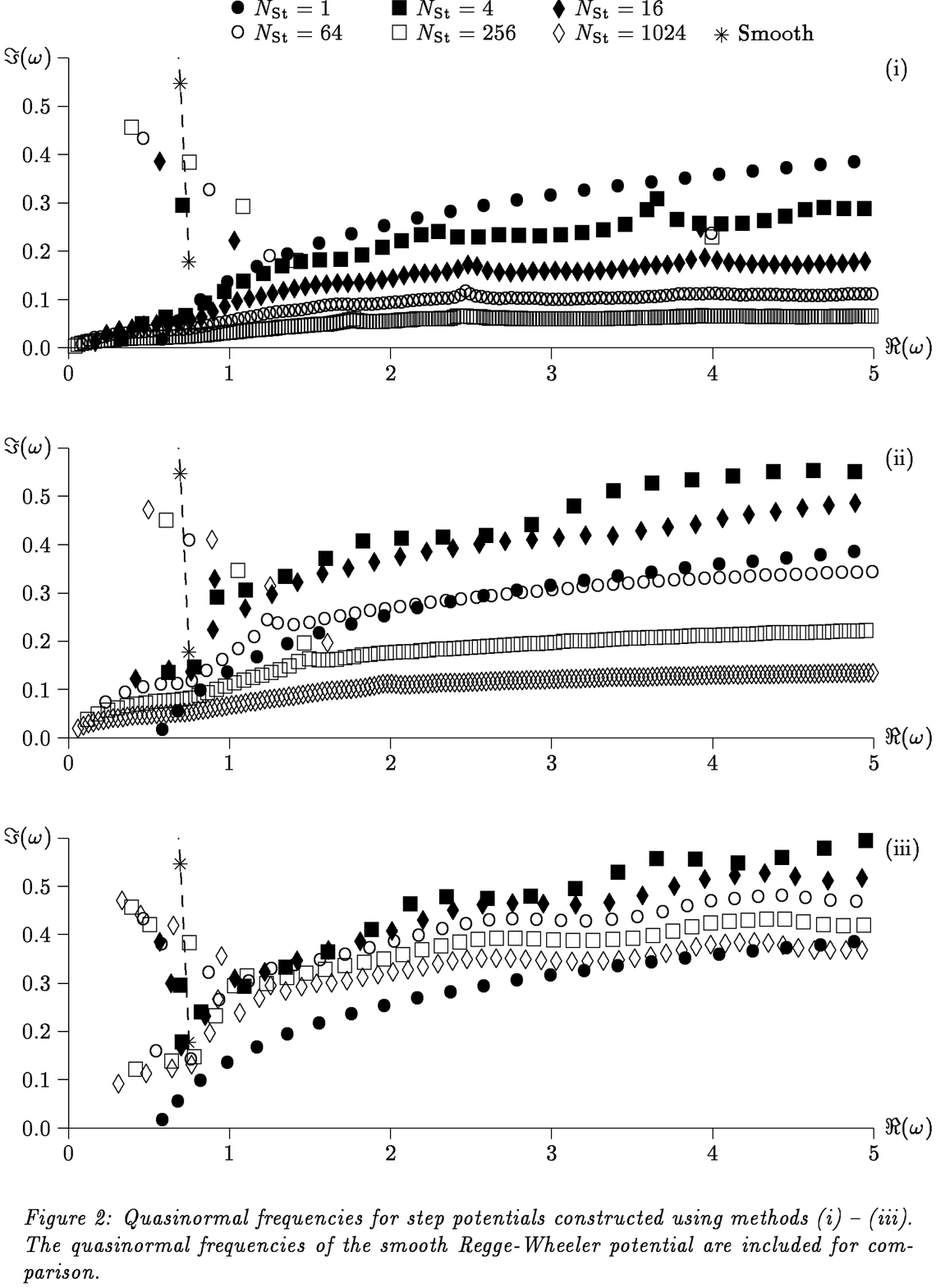}
\endinsert

\subtitle{III. Results}
The quasinormal frequencies determined in this way are shown in 
Fig.~2. The details of the behavior of the step potentials'
quasinormal frequencies, such as the value of the fundamental
frequency, the imaginary parts of the 'lined up' frequencies, and even
their spacing, depend on the method of construction.
The dominating features, however, are independent of it. In all cases, the
spacing between the frequencies becomes closer as the number of steps
increases, but they generally occupy the same part of the complex
plane. In particular, they remain 'lined up' more or less parallel to
the real axis, rather than to the imaginary axis as the quasinormal
frequencies of the smooth Regge-Wheeler potential. There is no
indication that the frequencies cease to reach arbitrarily large real
parts, i.e. arbitrarily small oscillation periods, as the number of
steps becomes very large.

As $\Nst$ increases, there are individual frequencies 'escaping'
from the line of frequencies towards the imaginary axis. However,
even these frequencies do not seem to approach the quasinormal
frequencies of the Regge-Wheeler potential.

%
%
%
%
%
%
\def\real#1{{#1\phantom{0.}}}
\def\imag#1{{\phantom{0.}#1}}
%
%
\def\qnfacvr{}
\def\qnfagvr{}
\def\qnfaivr{}
\def\qnfbcvr{}
\def\qnfbgvr{}
\def\qnfbkvr{}
\def\qnfccvr{}
\def\qnfcgvr{}
\def\qnfckvr{}
\def\qnfacvi{}
\def\qnfagvi{}
\def\qnfaivi{}
\def\qnfbcvi{}
\def\qnfbgvi{}
\def\qnfbkvi{}
\def\qnfccvi{}
\def\qnfcgvi{}
\def\qnfckvi{}
\def\qnfacwr{}
\def\qnfagwr{}
\def\qnfaiwr{}
\def\qnfbcwr{}
\def\qnfbgwr{}
\def\qnfbkwr{}
\def\qnfccwr{}
\def\qnfcgwr{}
\def\qnfckwr{}
\def\qnfacwi{}
\def\qnfagwi{}
\def\qnfaiwi{}
\def\qnfbcwi{}
\def\qnfbgwi{}
\def\qnfbkwi{}
\def\qnfccwi{}
\def\qnfcgwi{}
\def\qnfckwi{}
\def\qnfacxr{}
\def\qnfagxr{}
\def\qnfaixr{}
\def\qnfbcxr{}
\def\qnfbgxr{}
\def\qnfbkxr{}
\def\qnfccxr{}
\def\qnfcgxr{}
\def\qnfckxr{}
\def\qnfacxi{}
\def\qnfagxi{}
\def\qnfaixi{}
\def\qnfbcxi{}
\def\qnfbgxi{}
\def\qnfbkxi{}
\def\qnfccxi{}
\def\qnfcgxi{}
\def\qnfckxi{}
\def\qnfacyr{}
\def\qnfagyr{}
\def\qnfaiyr{}
\def\qnfbcyr{}
\def\qnfbgyr{}
\def\qnfbkyr{}
\def\qnfccyr{}
\def\qnfcgyr{}
\def\qnfckyr{}
\def\qnfacyi{}
\def\qnfagyi{}
\def\qnfaiyi{}
\def\qnfbcyi{}
\def\qnfbgyi{}
\def\qnfbkyi{}
\def\qnfccyi{}
\def\qnfcgyi{}
\def\qnfckyi{}
\def\qnfaczr{}
\def\qnfagzr{}
\def\qnfaizr{}
\def\qnfbczr{}
\def\qnfbgzr{}
\def\qnfbkzr{}
\def\qnfcczr{}
\def\qnfcgzr{}
\def\qnfckzr{}
\def\qnfaczi{}
\def\qnfagzi{}
\def\qnfaizi{}
\def\qnfbczi{}
\def\qnfbgzi{}
\def\qnfbkzi{}
\def\qnfcczi{}
\def\qnfcgzi{}
\def\qnfckzi{}
%
%
%
%
\def\qnfacar{\real{ 0.329}}
\def\qnfacai{\imag{ 0.021}}
\def\qnfacbr{\real{ 0.457}}
\def\qnfacbi{\imag{ 0.051}}
\def\qnfaccr{\real{ 0.604}}
\def\qnfacci{\imag{ 0.064}}
\def\qnfacdr{\real{ 0.727}}
\def\qnfacdi{\imag{ 0.068}}
\def\qnfacer{\real{ 0.847}}
\def\qnfacei{\imag{ 0.093}}
\def\qnfacfr{\real{ 1.443}}
\def\qnfacfi{\imag{ 0.180}}
\def\qnfacgr{\real{ 2.673}}
\def\qnfacgi{\imag{ 0.234}}
\def\qnfachr{\real{ 6.460}}
\def\qnfachi{\imag{ 0.323}}
\def\qnfacir{\real{12.758}}
\def\qnfacii{\imag{ 0.365}}
\def\qnfacvr{\real{ 0.707}}
\def\qnfacvi{\imag{ 0.295}}
%
%
%
\def\qnfagar{\real{ 0.084}}
\def\qnfagai{\imag{ 0.006}}
\def\qnfagbr{\real{ 0.120}}
\def\qnfagbi{\imag{ 0.015}}
\def\qnfagcr{\real{ 0.162}}
\def\qnfagci{\imag{ 0.020}}
\def\qnfagdr{\real{ 0.202}}
\def\qnfagdi{\imag{ 0.022}}
\def\qnfager{\real{ 0.241}}
\def\qnfagei{\imag{ 0.025}}
\def\qnfagfr{\real{ 0.441}}
\def\qnfagfi{\imag{ 0.031}}
\def\qnfaggr{\real{ 0.824}}
\def\qnfaggi{\imag{ 0.043}}
\def\qnfaghr{\real{ 1.957}}
\def\qnfaghi{\imag{ 0.091}}
\def\qnfagir{\real{ 3.788}}
\def\qnfagii{\imag{ 0.110}}
\def\qnfagvr{\real{ 1.250}}
\def\qnfagvi{\imag{ 0.189}}
\def\qnfagwr{\real{ 0.875}}
\def\qnfagwi{\imag{ 0.327}}
\def\qnfagxr{\real{ 0.465}}
\def\qnfagxi{\imag{ 0.432}}
\def\qnfagyr{\real{ 3.991}}
\def\qnfagyi{\imag{ 0.236}}
%
%
\def\qnfaiar{\real{ 0.042}}
\def\qnfaiai{\imag{ 0.003}}
\def\qnfaibr{\real{ 0.061}}
\def\qnfaibi{\imag{ 0.008}}
\def\qnfaicr{\real{ 0.082}}
\def\qnfaici{\imag{ 0.010}}
\def\qnfaidr{\real{ 0.102}}
\def\qnfaidi{\imag{ 0.011}}
\def\qnfaier{\real{ 0.122}}
\def\qnfaiei{\imag{ 0.013}}
\def\qnfaifr{\real{ 0.225}}
\def\qnfaifi{\imag{ 0.016}}
\def\qnfaigr{\real{ 0.427}}
\def\qnfaigi{\imag{ 0.021}}
\def\qnfaihr{\real{ 1.023}}
\def\qnfaihi{\imag{ 0.033}}
\def\qnfaiir{\real{ 2.027}}
\def\qnfaiii{\imag{ 0.057}}
\def\qnfaivr{\real{ 1.415}}
\def\qnfaivi{\imag{ 0.182}}
\def\qnfaiwr{\real{ 1.084}}
\def\qnfaiwi{\imag{ 0.294}}
\def\qnfaixr{\real{ 0.751}}
\def\qnfaixi{\imag{ 0.385}}
\def\qnfaiyr{\real{ 0.391}}
\def\qnfaiyi{\imag{ 0.457}}
\def\qnfaizr{\real{ 3.997}}
\def\qnfaizi{\imag{ 0.229}}
%
%
\def\qnfbcar{\real{ 0.621}}
\def\qnfbcai{\imag{ 0.136}}
\def\qnfbcbr{\real{ 0.782}}
\def\qnfbcbi{\imag{ 0.147}}
\def\qnfbccr{\real{ 0.922}}
\def\qnfbcci{\imag{ 0.292}}
\def\qnfbcdr{\real{ 1.101}}
\def\qnfbcdi{\imag{ 0.306}}
\def\qnfbcer{\real{ 1.346}}
\def\qnfbcei{\imag{ 0.335}}
\def\qnfbcfr{\real{ 2.594}}
\def\qnfbcfi{\imag{ 0.419}}
\def\qnfbcgr{\real{ 5.150}}
\def\qnfbcgi{\imag{ 0.550}}
\def\qnfbchr{\real{12.862}}
\def\qnfbchi{\imag{ 0.710}}
\def\qnfbcir{\real{25.720}}
\def\qnfbcii{\imag{ 0.829}}
%
%
%
\def\qnfbgar{\real{ 0.229}}
\def\qnfbgai{\imag{ 0.073}}
\def\qnfbgbr{\real{ 0.352}}
\def\qnfbgbi{\imag{ 0.093}}
\def\qnfbgcr{\real{ 0.464}}
\def\qnfbgci{\imag{ 0.104}}
\def\qnfbgdr{\real{ 0.574}}
\def\qnfbgdi{\imag{ 0.110}}
\def\qnfbger{\real{ 0.677}}
\def\qnfbgei{\imag{ 0.112}}
\def\qnfbgfr{\real{ 1.154}}
\def\qnfbgfi{\imag{ 0.208}}
\def\qnfbggr{\real{ 2.060}}
\def\qnfbggi{\imag{ 0.271}}
\def\qnfbghr{\real{ 4.897}}
\def\qnfbghi{\imag{ 0.342}}
\def\qnfbgir{\real{ 9.684}}
\def\qnfbgii{\imag{ 0.392}}
\def\qnfbgvr{\real{ 0.749}}
\def\qnfbgvi{\imag{ 0.408}}
%
%
\def\qnfbkar{\real{ 0.060}}
\def\qnfbkai{\imag{ 0.020}}
\def\qnfbkbr{\real{ 0.093}}
\def\qnfbkbi{\imag{ 0.026}}
\def\qnfbkcr{\real{ 0.124}}
\def\qnfbkci{\imag{ 0.030}}
\def\qnfbkdr{\real{ 0.156}}
\def\qnfbkdi{\imag{ 0.033}}
\def\qnfbker{\real{ 0.187}}
\def\qnfbkei{\imag{ 0.035}}
\def\qnfbkfr{\real{ 0.338}}
\def\qnfbkfi{\imag{ 0.043}}
\def\qnfbkgr{\real{ 0.635}}
\def\qnfbkgi{\imag{ 0.051}}
\def\qnfbkhr{\real{ 1.517}}
\def\qnfbkhi{\imag{ 0.093}}
\def\qnfbkir{\real{ 2.874}}
\def\qnfbkii{\imag{ 0.120}}
\def\qnfbkvr{\real{ 1.610}}
\def\qnfbkvi{\imag{ 0.197}}
\def\qnfbkwr{\real{ 1.254}}
\def\qnfbkwi{\imag{ 0.315}}
\def\qnfbkxr{\real{ 0.893}}
\def\qnfbkxi{\imag{ 0.411}}
\def\qnfbkyr{\real{ 0.496}}
\def\qnfbkyi{\imag{ 0.473}}
%
%
\def\qnfccar{\real{ 0.696}}
\def\qnfccai{\imag{ 0.180}}
\def\qnfccbr{\real{ 0.695}}
\def\qnfccbi{\imag{ 0.292}}
\def\qnfcccr{\real{ 0.816}}
\def\qnfccci{\imag{ 0.232}}
\def\qnfccdr{\real{ 1.079}}
\def\qnfccdi{\imag{ 0.287}}
\def\qnfccer{\real{ 1.335}}
\def\qnfccei{\imag{ 0.326}}
\def\qnfccfr{\real{ 2.569}}
\def\qnfccfi{\imag{ 0.468}}
\def\qnfccgr{\real{ 5.133}}
\def\qnfccgi{\imag{ 0.599}}
\def\qnfcchr{\real{12.808}}
\def\qnfcchi{\imag{ 0.737}}
\def\qnfccir{\real{25.623}}
\def\qnfccii{\imag{ 0.855}}
%
%
\def\qnfcgar{\real{ 0.756}}
\def\qnfcgai{\imag{ 0.140}}
\def\qnfcgbr{\real{ 0.533}}
\def\qnfcgbi{\imag{ 0.157}}
\def\qnfcgcr{\real{ 0.924}}
\def\qnfcgci{\imag{ 0.259}}
\def\qnfcgdr{\real{ 1.111}}
\def\qnfcgdi{\imag{ 0.298}}
\def\qnfcger{\real{ 1.253}}
\def\qnfcgei{\imag{ 0.325}}
\def\qnfcgfr{\real{ 2.007}}
\def\qnfcgfi{\imag{ 0.380}}
\def\qnfcggr{\real{ 3.502}}
\def\qnfcggi{\imag{ 0.431}}
\def\qnfcghr{\real{ 8.032}}
\def\qnfcghi{\imag{ 0.527}}
\def\qnfcgir{\real{15.602}}
\def\qnfcgii{\imag{ 0.583}}
\def\qnfcgvr{\real{ 0.872}}
\def\qnfcgvi{\imag{ 0.320}}
\def\qnfcgwr{\real{ 0.578}}
\def\qnfcgwi{\imag{ 0.376}}
\def\qnfcgxr{\real{ 0.465}}
\def\qnfcgxi{\imag{ 0.432}}
%
%
\def\qnfckar{\real{ 0.311}}
\def\qnfckai{\imag{ 0.091}}
\def\qnfckbr{\real{ 0.485}}
\def\qnfckbi{\imag{ 0.113}}
\def\qnfckcr{\real{ 0.643}}
\def\qnfckci{\imag{ 0.123}}
\def\qnfckdr{\real{ 0.766}}
\def\qnfckdi{\imag{ 0.131}}
\def\qnfcker{\real{ 0.879}}
\def\qnfckei{\imag{ 0.197}}
\def\qnfckfr{\real{ 1.351}}
\def\qnfckfi{\imag{ 0.284}}
\def\qnfckgr{\real{ 2.296}}
\def\qnfckgi{\imag{ 0.340}}
\def\qnfckhr{\real{ 5.125}}
\def\qnfckhi{\imag{ 0.377}}
\def\qnfckir{\real{ 9.837}}
\def\qnfckii{\imag{ 0.424}}
\def\qnfckvr{\real{ 0.951}}
\def\qnfckvi{\imag{ 0.355}}
\def\qnfckwr{\real{ 0.652}}
\def\qnfckwi{\imag{ 0.419}}
\def\qnfckxr{\real{ 0.450}}
\def\qnfckxi{\imag{ 0.442}}
\def\qnfckyr{\real{ 0.335}}
\def\qnfckyi{\imag{ 0.471}}
\midinsert
\centerline{\vbox{\eightrm
  \offinterlineskip \tabskip=0pt
  \halign{\strut
          \vrule#&              
          \hfil $ #~$ &\vrule#& 
          \hfil $\,#$ &\vrule#& 
          \hfil $\,#$ &\vrule#& 
          \hfil $\,#$ &\vrule#& 
          \hfil $\,#$ &\vrule#& 
          \hfil $\,#$ &\vrule#& 
          \hfil $\,#$ &\vrule#& 
          \hfil $\,#$ &\vrule#& 
          \hfil $\,#$ &\vrule#& 
          \hfil $\,#$ &\vrule#& 
          \hfil $\,#$ &\vrule#  
          \cr
     \noalign{\hrule}
     & \hfill && \multispan 5 \hfil (i)   Variable step size \hfil &&
                 \multispan 5 \hfil (ii)  Fixed step size    \hfil &&
                 \multispan 5 \hfil (iii) Exponential cutoff \hfil &&
                 \multispan 1 \hfil smooth             \hfil &          \cr
     \noalign{\hrule}
 & \omit ~Index && \omit ~4 steps  &&  64~~   &&  256~  &&
                   \omit ~4 steps  &&  64~~   && 1024~  &&
                   \omit ~4 steps  &&  64~~   && 1024~  &&        &  \cr
     \noalign{\hrule}
    &  1&& \qnfacar  &&  \qnfagar  && \qnfaiar &&
           \qnfbcar  &&  \qnfbgar  && \qnfbkar &&
           \qnfccar  &&  \qnfcgar  && \qnfckar &&          &  \cr
    &   && \qnfacai  &&  \qnfagai  && \qnfaiai &&
           \qnfbcai  &&  \qnfbgai  && \qnfbkai &&
           \qnfccai  &&  \qnfcgai  && \qnfckai &&          &  \cr
     \noalign{\hrule}
    &  2&& \qnfacbr  &&  \qnfagbr  && \qnfaibr &&
           \qnfbcbr  &&  \qnfbgbr  && \qnfbkbr &&
           \qnfccbr  &&  \qnfcgbr  && \qnfckbr &&          &  \cr
    &   && \qnfacbi  &&  \qnfagbi  && \qnfaibi &&
           \qnfbcbi  &&  \qnfbgbi  && \qnfbkbi &&
           \qnfccbi  &&  \qnfcgbi  && \qnfckbi &&          &  \cr
     \noalign{\hrule}
    &  3&& \qnfaccr  &&  \qnfagcr  && \qnfaicr &&
           \qnfbccr  &&  \qnfbgcr  && \qnfbkcr &&
           \qnfcccr  &&  \qnfcgcr  && \qnfckcr &&          &  \cr
    &   && \qnfacci  &&  \qnfagci  && \qnfaici &&
           \qnfbcci  &&  \qnfbgci  && \qnfbkci &&
           \qnfccci  &&  \qnfcgci  && \qnfckci &&          &  \cr
     \noalign{\hrule}
    &  4&& \qnfacdr  &&  \qnfagdr  && \qnfaidr &&
           \qnfbcdr  &&  \qnfbgdr  && \qnfbkdr &&
           \qnfccdr  &&  \qnfcgdr  && \qnfckdr &&          &  \cr
    &   && \qnfacdi  &&  \qnfagdi  && \qnfaidi &&
           \qnfbcdi  &&  \qnfbgdi  && \qnfbkdi &&
           \qnfccdi  &&  \qnfcgdi  && \qnfckdi &&          &  \cr
     \noalign{\hrule}
    &  5&& \qnfacer  &&  \qnfager  && \qnfaier &&
           \qnfbcer  &&  \qnfbger  && \qnfbker &&
           \qnfccer  &&  \qnfcger  && \qnfcker &&          &  \cr
    &   && \qnfacei  &&  \qnfagei  && \qnfaiei &&
           \qnfbcei  &&  \qnfbgei  && \qnfbkei &&
           \qnfccei  &&  \qnfcgei  && \qnfckei &&          &  \cr
     \noalign{\hrule}
    & 10&& \qnfacfr  &&  \qnfagfr  && \qnfaifr &&
           \qnfbcfr  &&  \qnfbgfr  && \qnfbkfr &&
           \qnfccfr  &&  \qnfcgfr  && \qnfckfr &&          &  \cr
    &   && \qnfacfi  &&  \qnfagfi  && \qnfaifi &&
           \qnfbcfi  &&  \qnfbgfi  && \qnfbkfi &&
           \qnfccfi  &&  \qnfcgfi  && \qnfckfi &&          &  \cr
     \noalign{\hrule}
    & 20&& \qnfacgr  &&  \qnfaggr  && \qnfaigr &&
           \qnfbcgr  &&  \qnfbggr  && \qnfbkgr &&
           \qnfccgr  &&  \qnfcggr  && \qnfckgr &&          &  \cr
    &   && \qnfacgi  &&  \qnfaggi  && \qnfaigi &&
           \qnfbcgi  &&  \qnfbggi  && \qnfbkgi &&
           \qnfccgi  &&  \qnfcggi  && \qnfckgi &&          &  \cr
     \noalign{\hrule}
    & 50&& \qnfachr  &&  \qnfaghr  && \qnfaihr &&
           \qnfbchr  &&  \qnfbghr  && \qnfbkhr &&
           \qnfcchr  &&  \qnfcghr  && \qnfckhr &&          &  \cr
    &   && \qnfachi  &&  \qnfaghi  && \qnfaihi &&
           \qnfbchi  &&  \qnfbghi  && \qnfbkhi &&
           \qnfcchi  &&  \qnfcghi  && \qnfckhi &&          &  \cr
     \noalign{\hrule}
    &100&& \qnfacir  &&  \qnfagir  && \qnfaiir &&
           \qnfbcir  &&  \qnfbgir  && \qnfbkir &&
           \qnfccir  &&  \qnfcgir  && \qnfckir &&          &  \cr
    &   && \qnfacii  &&  \qnfagii  && \qnfaiii &&
           \qnfbcii  &&  \qnfbgii  && \qnfbkii &&
           \qnfccii  &&  \qnfcgii  && \qnfckii &&          &  \cr
     \noalign{\hrule}
    & o1&& \qnfacvr  &&  \qnfagvr  && \qnfaivr &&
           \qnfbcvr  &&  \qnfbgvr  && \qnfbkvr &&
           \qnfccvr  &&  \qnfcgvr  && \qnfckvr && \real{0.747} &  \cr
    &   && \qnfacvi  &&  \qnfagvi  && \qnfaivi &&
           \qnfbcvi  &&  \qnfbgvi  && \qnfbkvi &&
           \qnfccvi  &&  \qnfcgvi  && \qnfckvi && \imag{0.178} &  \cr
     \noalign{\hrule}
    & o2&& \qnfacwr  &&  \qnfagwr  && \qnfaiwr &&
           \qnfbcwr  &&  \qnfbgwr  && \qnfbkwr &&
           \qnfccwr  &&  \qnfcgwr  && \qnfckwr && \real{0.693} &  \cr
    &   && \qnfacwi  &&  \qnfagwi  && \qnfaiwi &&
           \qnfbcwi  &&  \qnfbgwi  && \qnfbkwi &&
           \qnfccwi  &&  \qnfcgwi  && \qnfckwi && \imag{0.548} &  \cr
     \noalign{\hrule}
    & o3&& \qnfacxr  &&  \qnfagxr  && \qnfaixr &&
           \qnfbcxr  &&  \qnfbgxr  && \qnfbkxr &&
           \qnfccxr  &&  \qnfcgxr  && \qnfckxr && \real{0.602} &  \cr
    &   && \qnfacxi  &&  \qnfagxi  && \qnfaixi &&
           \qnfbcxi  &&  \qnfbgxi  && \qnfbkxi &&
           \qnfccxi  &&  \qnfcgxi  && \qnfckxi && \imag{0.957} &  \cr
     \noalign{\hrule}
    & o4&& \qnfacyr  &&  \qnfagyr  && \qnfaiyr &&
           \qnfbcyr  &&  \qnfbgyr  && \qnfbkyr &&
           \qnfccyr  &&  \qnfcgyr  && \qnfckyr &&          &  \cr
    &   && \qnfacyi  &&  \qnfagyi  && \qnfaiyi &&
           \qnfbcyi  &&  \qnfbgyi  && \qnfbkyi &&
           \qnfccyi  &&  \qnfcgyi  && \qnfckyi &&          &  \cr
     \noalign{\hrule}
    & o5&& \qnfaczr  &&  \qnfagzr  && \qnfaizr &&
           \qnfbczr  &&  \qnfbgzr  && \qnfbkzr &&
           \qnfcczr  &&  \qnfcgzr  && \qnfckzr &&          &  \cr
    &   && \qnfaczi  &&  \qnfagzi  && \qnfaizi &&
           \qnfbczi  &&  \qnfbgzi  && \qnfbkzi &&
           \qnfcczi  &&  \qnfcgzi  && \qnfckzi &&          &  \cr
     \noalign{\hrule}
                                                             }}}
\smallskip
{\it Table 1: Quasinormal frequencies of step potentials for all three
methods of construction and various step sizes. Each entry lists the
real part of the frequency as the upper left and the imaginary part
as the lower right number. The first three frequencies of the smooth
Regge-Wheeler potential are included for comparison.
}
\endinsert

Table 1 lists some quasinormal frequencies for all three
methods of construction and different numbers of steps, along with the
'outlying' frequencies and the fundamental quasinormal frequencies of
the smooth Regge-Wheeler potential. In the case of method (i), the
long step at large $x$ causes numerical instabilities. Therefore,
results for $\Nst = 1024$ are not available; the frequencies
for $\Nst = 256$ are listed instead.

In order to check whether the behavior we found could be the result of some
numerical artifact, we have explicitly calculated the time evolution
of some Cauchy data subject to the time dependent version of Eq.~(\EqSchr)
\def\Eqtime{8}
$$ {\partial^2 \psi \over \partial x^2}
      - {\partial^2 \psi \over \partial t^2} - V(x) \psi = 0 ,
\eqno{(\Eqtime)} $$
inserting step potentials constructed using method (i), with 
$\Nst = 1$, $\Nst = 4$, and $\Nst = 64$, for $V(x)$. 
The initial data consisted of a bell shaped peak with compact support,
incident onto the
potential from the left, the point of observation is at $x = 120$. 
The result for $\Nst = 4$ is shown in Fig.~3, and it manages to combine two
seemingly contradictory requirements: It is only slightly different
from the time evolution data for the continuous Regge-Wheeler
potential (included in
Fig.~3 for comparison), and it also represents the totally different
spectrum of quasinormal frequencies of the step potential. The
similarity is seen shortly after the initial burst has passed (Fig.~3(i));
actually, the time evolution here is dominated by the fundamental quasinormal
frequency of the {\it smooth} potential, which is not present in the
spectrum of the step potential. At very late times, however, the
perturbation oscillates with the fundamental frequency of the {\it step }
potential (Fig.~3(ii)), while the data for the smooth potential is damped 
more strongly and is therefore invisible in Fig.~3(ii). However, this
difference concerns only a tiny fraction of the total energy radiated
by the system.

%
%
\midinsert
\centerline{%
\vbox{
  \offinterlineskip \tabskip=0pt
  \halign{\strut
          \vrule#&              
          \hfil $ #~$ &\vrule#& 
          \hfil $\,#$ &         
        ~ \hfil $#$ &\vrule#&   
          \hfil $\,#$ &         
        ~ \hfil $#$ &\vrule#&   
          \hfil $\,#$ &         
        ~ \hfil $#$ &\vrule#&   
          \hfil $\,#$ &         
        ~ \hfil $#$ &\vrule#    
          \cr
     \noalign{\hrule}
     & \hfill && \multispan 5 \hfil Stationary calculation \hfil &&
                 \multispan 5 \hfil Time evolution         \hfil &  \cr
     \noalign{\hrule}
     & \hfill && \multispan 2 \hfil  1 step  \hfil &&
                 \multispan 2 \hfil 64 steps \hfil &&
                 \multispan 2 \hfil  1 step  \hfil &&
                 \multispan 2 \hfil 64 steps \hfil &     \cr
     \noalign{\hrule}
 & \omit ~Index 
       && \omit \hfil Re($\omega$) \hfil & \omit \hfil Im($\omega$) \hfil &&
          \omit \hfil Re($\omega$) \hfil & \omit \hfil Im($\omega$) \hfil &&
          \omit \hfil Re($\omega$) \hfil & \omit \hfil Im($\omega$) \hfil &&
          \omit \hfil Re($\omega$) \hfil & \omit \hfil Im($\omega$) \hfil & \cr
     \noalign{\hrule}
    &  1&& 0.5836    &  0.0165     &&
           0.0841    &  0.0064     &&
           0.5828    &  0.0169     &&
           0.0841    &  0.0064     &                \cr
    &  2&& 0.6786    &  0.0550     &&
           0.1203    &  0.0151     &&
           0.6792    &  0.0566     &&
           0.1203    &  0.0152     &                \cr
    &  3&& 0.8187    &  0.0975     &&
                     &             &&
           0.8245    &  0.1022     && 
                     &             &                \cr
    &  4&& 0.9859    &  0.1351     &&
                     &             &&
           1.0333    &  0.1374     &&
                     &             &                \cr
     \noalign{\hrule}
                                                             }}}
\smallskip
{\it Table 2: Comparison of quasinormal frequencies obtained from
  the stationary calculation described in this paper with frequencies
  obtained by a fit of time evolution data.
}
\endinsert

\topinsert
\epsfxsize=\hsize \epsfbox{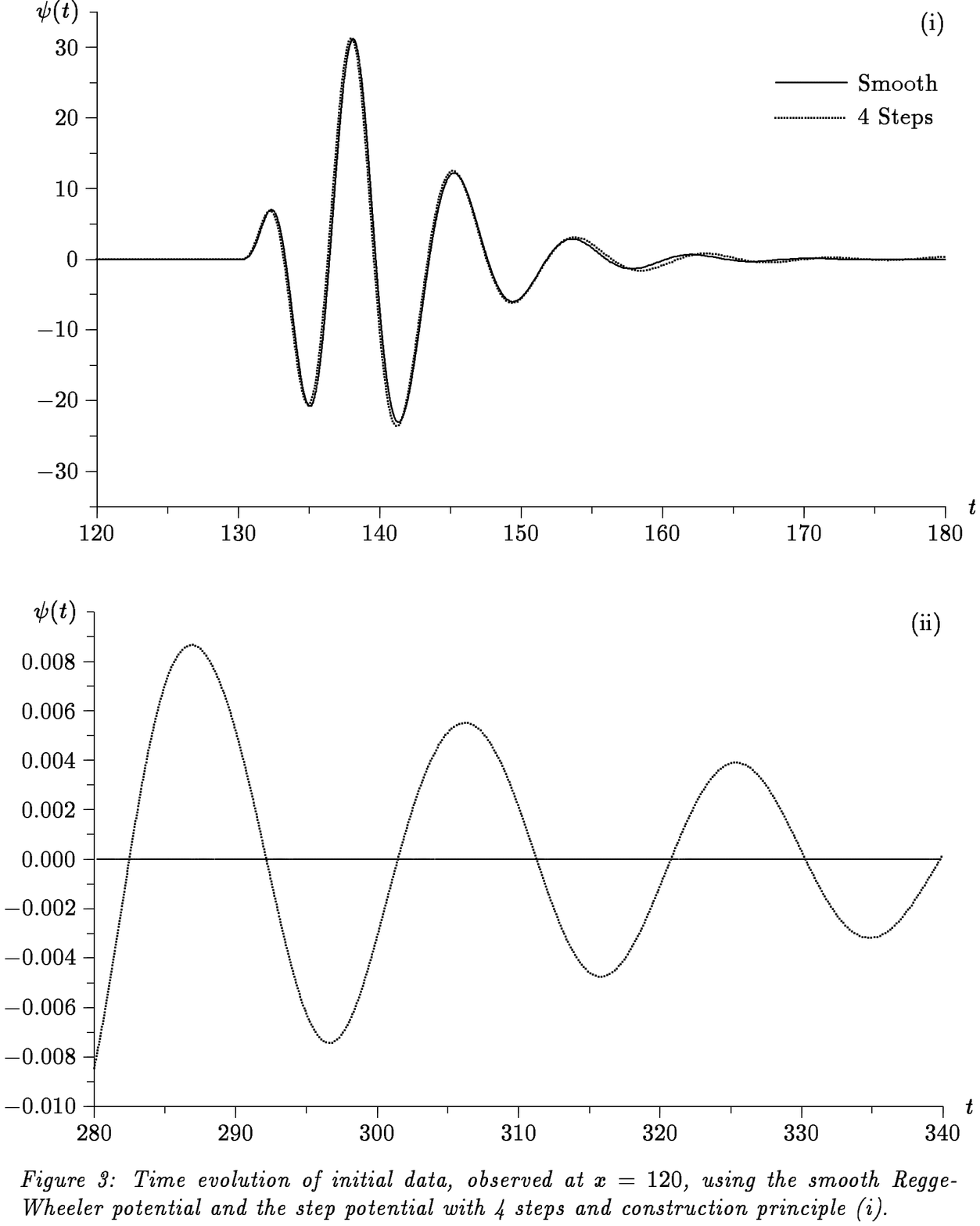}
\endinsert

We determined the dominating frequencies of these late time 
oscillations by fitting quasinormal modes to the time dependent
solution, their frequencies not being fixed, but parameters of the fit
procedure. The fundamental frequencies obtained in this way are
given in Tab.~2, they agree very well with the fundamental
quasinormal frequencies obtained from the stationary calculation.
Some deviation is expected, since the time dependent calculation works
on a fixed grid and does not know about the exact location of the jumps in
the step potential, or even about the potential being a step
potential at all.

\subtitle{IV. Discussion}
Introducing discontinuities in the Regge-Wheeler potential has a significant
influence on the quasinormal mode spectrum, even if the jumps 
become very small. Most of the frequencies become 'lined up' roughly
parallel to the real axis, as they do for the simple square barrier
potential, rather than being lined up along the imaginary axis like
the quasinormal frequencies of the smooth Regge-Wheeler potential.
This is a necessary, but not sufficient, condition for the quasinormal
modes forming a complete set for the solutions of the perturbation
equation.

Time evolution computations confirm that the very late time behavior
of some initial perturbation is indeed dominated by the fundamental 
quasinormal mode of the step potential under consideration, rather
than by the fundamental mode or the tail belonging to the smooth 
potential. However, due to the exponential damping, 
this very late time behavior
represents only a tiny fraction of the total response of a black
hole, or some other system, to an initial perturbation.
Most of the total response occurs at earlier times, where it is
dominated by quasinormal ringing at the fundamental frequency of the
smooth potential.

Our results suggest that it is indeed possible to obtain an expanded
set of quasinormal modes which can completely represent the time
evolution of perturbations of black holes. This conclusion increases 
the significance of quasinormal modes for the study of perturbations
of black holes. 
On the other hand, we also lose something: None of
the frequencies of the new modes is obvious in the largest part of the
time evolution of the perturbation. It is generally assumed that at
least the fundamental quasinormal frequencies have physical meaning in
the sense that they will dominate the time evolution of a perturbation
of the black hole, as it is known for the quasinormal frequencies
of the continuous Regge-Wheeler potential. The quasinormal spectrum of
a very similar step potential, on the other hand, might actually lead
to a complete system of quasinormal modes, but there is no single mode
or frequency which has an obvious relationship to the time dependence
of the perturbation. 

\def\LitPT{12}
We have also seen that even a very small deviation from the original
potential leads to a completely different quasinormal mode spectrum.
It appears rather farfetched to assign significance to quantities
which are so sensitive to small details, rather than to the overall
features of the problem. On the other hand, this sensitivity might
well be a consequence of some specific property of the step
potentials, as it does not seem to occur for other cases that have
been studied: Ferrari and Mashhoon [\LitPT], for example, have replaced the
Regge-Wheeler potential by a Poeschl-Teller potential, approximating
the Regge-Wheeler potential mainly around its maximum. They achieve
good agreement between the fundamental quasinormal frequencies of the
two potentials.

How, then, can we know if the quasinormal spectrum of
any given system does indeed tell us something about the time evolution
of a perturbation of this systems? What are the criteria which
distinguish a ``meaningful'' from a ``meaningless'' spectrum? Are
quasinormal frequencies always strongly sensitive to small changes in
some parameters of the problem? And can
we find another way to obtain an ``expanded'' set of quasinormal modes
which is complete, but leaves the ``original'' modes more or less
intact? 

There is an intriguing parallel between the quasinormal spectrum we
have found for step versions of the Regge-Wheeler potential and those
of neutron stars: The frequencies of the so-called wave modes, which
are associated mainly with the metric outside the star, rather than
with the fluid inside, show a picture rather similar
to what we have found here for the quasinormal frequencies of the step
potentials, with the frequencies of the $w$-modes corresponding to the
frequencies lined up along the real axis, and the frequencies of the
\wII-modes corresponding to the 'outlying' frequencies~[\Litwii]. A better
understanding of neutron star perturbations as well as Schr\"odinger or
wave equations with discontinuities is required to decide whether this
parallel is a coincidence or not. It is possible, however, that the
junction conditions at the surface of the neutron star provide an
analogy for some discontinuity in a potential.

The present work shows that the significance  of the quasinormal mode
spectrum of a black hole is not yet clearly understood. 
The most intriguing questions may  be whether it is possible to
``complete'' the set of quasinormal modes in such a way that the
fundamental modes still represent the major features of the time
evolution, and 
how the excitation of a particular quasinormal mode can be measured,
or even be given a unanimous meaning.

\vbox{

\subtitle{Acknowledgments}
We wish to thank Richard Price for bringing the completeness problem
to our attention, and for many enlightening discussions. Jorge Pullin made
available a code which helped with the computation of the time
evolution data.
}

\vfill

\subtitle{References}
\parindent=2em%
\item{[\Litw]}        K. D. Kokkotas and B. F. Schutz, 
                      Mon. Not. R. Astron. Soc. {\bf 255}, 119 (1992)
\item{[\Litwii]}      M. Leins, H.-P. Nollert, and M. H. Soffel,
                      Phys. Rev. D {\bf 48}, 3467 (1993)
\item{[\Litqnmoc]}    P. T. Leung, S. Y. Liu, S. S. Tong, and K. Young,
                      Phys. Rev. A {\bf 49}, 3068 (1994)
\item{[\Litcompoc]}   P. T. Leung, S. Y. Liu, and K. Young,
                      Phys. Rev. A {\bf 49}, 3057 (1994)
\item{[\Litqnmbh]}    W. H. Press, Ap. J. {\bf 170}, L105 (1971)
\item{[\Litnumqnm]}   C. V. Vishveshwara, Nature {\bf 227}, 936 (1970)
\item{[\LitRWeq]}     S. Chandrasekhar and S. Detweiler, 
                      Proc. R. Soc. London A {\bf 344}, 441 (1975)
\item{[\LitInfMany]}  A. Bachelot and A. Motet-Bachelot, Proceedings
                      of the ``IV International Conference on Hyperbolic
                      Problems'', Taosmina, ed. Vieweg (1992)
\item{[\LitTails]}    E. Leaver, Phys. Rev. D {\bf 34}, 384 (1986), 
                         Phys. Rev. D {\bf 38}, 725 (erratum)
\item{[\LitTorsFib]}  R. H. Price and V. Husain, Phys. Rev. Lett. 
                       {\bf 68}, 1973 (1992)
\item{[\Litcomppot]}  E. S. C. Ching, P. T. Leung, W. M. Suen, and 
                      K. Young, Phys. Rev. Lett. {\bf 74}, 4588
                      (1995), gr-qc/9408043
\item{[\LitPT]}    V. Ferrari and B. Mashhoon, Phys. Rev. Lett. 
                   {\bf 52}, 1361 (1984) \hfill\break
                   V. Ferrari and B. Mashhoon, Phys. Rev. D {\bf 30},
                   295 (1984)

\end